\documentstyle[twocolumn,prl,aps,epsfig]{revtex}
\tightenlines

\begin{document}
\draft
\def\be{\begin{equation}}
\def\ee{\end{equation}}
\def\bfi{\begin{figure}}
\def\efi{\end{figure}}
\def\bea{\begin{eqnarray}}
\def\eea{\end{eqnarray}}
\def\vp{\varphi}
\title{Comment }

{\bf Comment on ``Aging, phase ordering and conformal invariance''}

In a recent letter\cite{HPGL} Henkel, Pleimling, Godr\`eche and Luck (HPGL) 
have derived the linear
response function of aging systems in the scaling regime of a quench at 
$T_C$ or below $T_C$  obtaining
$R(t,s) \sim s^{-1-a} x^{1+a-\lambda/z}(x-1)^{-1-a}$
where $z$ and $\lambda$ are the dynamic and the Fisher and Huse\cite{Fisher}
exponent, $a$ is the {\it response} exponent and $x=t/s$. 
While $z$ and $\lambda$ are
well known for most systems, relatively little is known about $a$. Here we
will restrict to the quench below $T_C$. Scaling of
$R(t,s)$ \cite{nota} implies scaling of both the zero 
field cooled magnetization (ZFCM)
$\chi(t,t_w) = \int_{t_w}^t ds R(t,s) \sim t_w^{-a_{\chi}}F(t/t_w)$ 
and of the thermoremanent magnetization (TRM) 
$\rho(t,t_w) = \int_0^{t_w} ds R(t,s) \sim  t_w^{-a_{\rho}}E(t/t_w)$
with $a_{\chi}=a_{\rho}=a$.
Assuming that scaling of the TRM holds in the time range of
their simulations
HPGL have measured $a_{\rho}$ in the ferromagnetic Ising model
obtaining $a_{\rho}=1/2$ for
$d=2,3$ (with logarithmic corrections for $d=2$). On the
other hand, putting together exact results for the $d=1$ Ising
model\cite{Ising1}, for the large $N$ model\cite{Ninfty} and
numerical simulations of the Ising model 
for $d=2,3,4$\cite{Ising} we have found a different result
for the ZFCM exponent
\be
a_{\chi} =  \left \{ \begin{array}{ll}
	\theta (d-d_L)/(d_U-d_L) & \mbox{for $d <d_U$ }      \\  
        \theta  & \mbox{for $d > d_U$}
                   \end{array}                                        
               \right .
\label{2}
\ee
with $\theta=1/2, d_L=1, d_U=3$ for Ising, $\theta=1, d_L=2, d_U=4$
for large $N$ and logarithmic corrections at $d_U$.
Therefore, concentrating on the Ising model, 
there is a large discrepancy for $d=2$ where HPGL find $a_{\rho}=1/2$
with logarithmic corrections as opposed to $a_{\chi}=1/4$ and a lesser one
for $d=3$ where
$a_{\rho}=1/2$ is opposed to  $a_{\chi}=1/2$ with logarithmic corrections.
It is the purpose of
this comment to show that the correct value of $a$ is 
given by $a_{\chi}$
and that the discrepancy with $a_{\rho}$ is due to a large preasymptotic 
correction affecting TRM in the range of $t_w$ explored by HPGL. 
In order to do this we 
have done over again the simulations of the Ising model in the same
conditions of HPGL, measuring both TRM and ZFCM over a wider
range of $t_w$. The scaling collapse of the ZFCM data
as $t_w$ is varied allows to identify $a_{\chi}$.
We find scaling with $a_{\chi}\simeq 0.27$ for $d=2$ and
$a_{\chi}\simeq 0.48$ with possible logarithmic corrections
for $d=3$ (Fig.1), consistently with Eq.~(\ref{2}). 

Next, in order to analyze the TRM data, let us
borrow from the large $N$ model\cite{Ninfty} the general form 
$R(t,s) = r_0[Y(s)/Y(t)](t-s)^{-1-a}$ which contains scaling taking 
$Y(t) \sim t^{-1 -a+\lambda/z}$ for times larger than a characteristic
microscopic time $t_0$. Then, for $t \gg t_w$
one has $\rho(t,t_w) = H(t_w)E(t/t_w)$ with $E(x) \sim x^{-\lambda/z}$.
About $H(t_w)$ not much can be said if $t_w <t_0$. Instead, if $t_w >t_0$
one has $H(t_w) \sim t_w^{-\lambda/z}[1+ (t_w/t^*)^{\lambda/z-a}]$ where
$t^*$ is the crossover time from $H(t_w) \sim t_w^{-\lambda/z}$ to
$H(t_w) \sim t_w^{-a}$. On the other hand, $a_{\rho}$ is defined by
$H(t_w) \sim t_w^{-a_{\rho}}$. Therefore, if $t^* > t_0$ there are three
regimes: 1) early regime for $t_w <t_0$ where $a_{\rho}$ is not well
defined,  2) preasymptotic regime $t_0 < t_w < t^*$ where 
$a_{\rho}= \lambda/z$, and 3) asymptotic regime $t_w > t^*$ where
$a_{\rho}=a$. Our data (Fig.1) show that even with the largest
$t_w$ we have reached the asymptotic
regime has not been entered, since we find an early regime 
followed by a power law with an exponent
$a_{\rho}$ which compares well with $\lambda/z$ ($\lambda /z=5/8 $ 
for $d=2$ and $\lambda/z=3/4$ for $d=3$).
The range of $t_w$ explored by
HPGL is in between the early and the preasymptotic regime where the slope of
$H(t_w)$ is compatible with an effective exponent close to $1/2$ both for
$d=2$ and $d=3$, but this is in no way related to the asymptotic value 
of $a_{\rho}$.

In conclusion, our results show that the measurement of the response 
exponent from TRM requires 
exceedingly large $t_w$, while it is manageable from ZFCM. We conjecture 
Eq.~(\ref{2}) to be of general validity for coarsening systems and 
work is under way to check it with conserved dynamics.

\begin{figure}
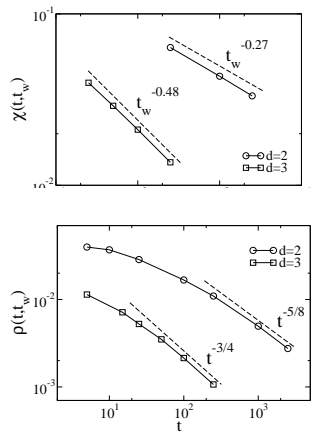

\noindent{\psfig{figure=figchi.eps,width=4cm,angle=0}}
\hspace{0.5 cm}{\psfig{figure=figrho.eps,width=4cm,angle=0}}
\caption{$\chi (t,t_w)$ (left) and $\rho (t,t_w)$ (right) are plotted 
against $t_w$ for fixed $t/t_w=10$. The same behavior is found for
different values of $t/t_w$.}
\end{figure}


\vspace{1cm}
F.Corberi, E.Lippiello and M.Zannetti

{\small Istituto Nazionale per la Fisica della Materia, Unit\`a di Salerno  
and Dipartimento di Fisica ``E.R.Caianiello'', Universit\`a di Salerno,
84081 Baronissi (Salerno), Italy}

\vspace{1cm}


\begin{thebibliography}{99}
\bibitem{nota}

Here we refer to the aging part of the response function,
see Ref.\cite{Ising}.
\end{thebibliography}
\end{document}